# Spatio-temporal programming of lyotropic phase transition in nanoporous microfluidic confinements

Vamseekrishna Ulaganathan[1] and Anupam Sengupta[1*]

[1]Physics of Living Matter Group, Department of Physics and Materials Science, University of Luxembourg, 162 A, Avenue de la Faencerie, L-1511, Luxembourg City, Luxembourg.

* To whom correspondence may be addressed: anupam.sengupta@uni.lu

**Funding:** This work is supported by the ATTRACT Investigator Grant (to A.S., Grant no. A17/MS/11572821/MBRACE) and CORE Grant (C19/MS/13719464/TOPOFLUME/Sengupta) of the Luxembourg National Research Fund; and donations from the CINVEN Foundation.

**Author Contributions:** A.S. conceptualized and developed the research plan, and directed all parts of the project. V.U. carried out experiments, analyzed the data and developed the mathematical model with inputs from A.S. V.U. and A.S. prepared the figures and wrote the paper.

**Competing interests:** Authors declare no competing interests.

**Data and materials availability:** All data and codes are available in the main text and the Supplementary materials, or can be obtained upon requests made to the corresponding author.

**Acknowledgements:** The authors thank Anshul Sharma, Sharadhi Nagaraja, Marco G. Mazza and Jayabrata Dhar for valuable discussions during the course of this work.

**Abstract**

Self-assembly of simple molecules into complex phases can be driven by physical constraints, for instance, due to selective molecular uptake by nanoporous surfaces. Despite the significance of surface-mediated assembly in the evolution of life, physical routes to molecular enrichment and assembly have remained overlooked. Here, using a lyotropic chromonic liquid crystal as model biological material, confined within nanoporous microfluidic environments, we study molecular assembly driven by nanoporous substrates. We demonstrate that nanoporous polydimethylsiloxane (PDMS) surfaces, due to selective permeation of water molecules, drive transition of disordered isotropic phase to ordered nematic, and higher order columnar phases under isothermal conditions. Synergistically, by tailoring the wettability, the surface-to-volume ratio, and surface topography of the confinements, we program the lyotropic phase transitions with a high degree of spatial and temporal control. Using a combination of timelapse polarized imaging, quantitative image processing, and a simple mathematical model, we analyze the phase transitions, and construct a master diagram capturing the role of surface wettability and channel geometry on programmable lyotropic phase transitions. Intrinsic PDMS nanoporosity and confinement cross-section, together with the imposed wettability regulate the rate of the N-M phase transition; whereas the microfluidic geometry and embedded topography enable phase transition at targeted locations. We harness the emergent long-range order during N-M transition to actuate elasto-advective transport of embedded micro-cargo, demonstrating particle manipulation concepts governed by tunable phase transitions. Our results present a programmable physical route to material assembly, and offer a new paradigm for assembling genetic components, biological cargo, and minimal synthetic cells.

**Main**

Microfluidics has enabled miniaturizing several technologies such chromatography, chemical reaction systems, micro mixers, micro-pumping systems, diffusive separation systems to name a few [1]. Soft lithography is one of methods employed in fabricating microfluidic devices using "soft" elastomers, such as polydimethylsiloxane (PDMS). PDMS is an organosilicon compound (a type of silicone oil) that forms an elastomer (a rubber-like material) when mixed with a cross-linker. Such property has allowed this material to be used in mainstream microfluidics as it can be cast onto the silicon wafers that contain channel design and then cured into a solid elastomer, with micron-sized details of channels imprinted. Such material is then plasma treated and bonded on the glass substrate, furnishing microchannels of varying degrees of complexity, which can be filled through cylindrical ports at either side, with desired material and visualized under a microscope (Figure 1a). The possibility of integrating topographical features [2], deformable walls [3], modular elements allowing tunable fields and gradients [4], and nanoporous elements [5] has enabled precise control and manipulation of multi-scale transport processes relevant for a wide range of applications in physical, chemical and biological settings.

PDMS, an inherently hydrophobic material, has a nanoporous ultrastructure [6]. Surface treatments including UV-ozone, and air or oxygen plasma exposure render PDMS hydrophilic [7 - 9], however with a steady hydrophobic recovery under normal environmental conditions [8, 9]. Plasma treatment oxidizes PDMS surface adding three to four oxygen atoms to a silicon atom [10], making it hydrophilic (Figure 1b). Dimethyloctadecyl[3-(trimethoxysilyl)propyl] ammonium chloride (DMOAP), a well-known silane used for incorporating hometropic (normal) molecular anchoring of liquid crystal molecules [11], induces hydrophobicity on both PDMS and glass surfaces. When LCs are introduced in the channel, the molecules align either parallel (perpendicular) to the plasma (DMOAP) treated surfaces [9].

The intrinsic material anisotropy of liquid crystals (LCs) makes LCs as complex functional material for microfluidics [12-14]. Harnessing the anisotropic coupling between the flow and the molecular ordering of thermotropic LCs (phase transition depends on temperature only) offers a novel paradigm to conventional microfluidic concepts based on isotropic fluids [15 - 18]. For instance, exploring LC matrix for enhanced particle transport in bulk [19], or tailoring topological defects as a system of self-assembled rails for micro-cargo transport [20] demonstrate unique attributes of LC-based microfluidic transport. The flow-driven orientational changes, along with the interplay of visco-elastic effects under microscale confinements and appropriate boundary conditions (surface anchoring), enable tunable flow profiles and surface-controlled hydrodynamic resistance [21, 22]. Numerous biological building blocks exhibit LC structures due to their ability to undergo reversible changes, both in morphology and in thermodynamic phase. Among others, biological LC structures include amphiphilic lipids, typically found in cellular membranes, the DNA in chromosomes, cytoskeletal and muscle proteins, and the collagens and proteoglycans of connective tissues [23]. Lyotropic LCs, formed due to the anisotropic assembly of water-soluble disc-shaped molecules, are biologically relevant owing to the weak interaction forces which underpin their self-assembly and high material response under external stimuli, including temperature, concentration, pH, and ionic content [24]. Lyotropic chromonic liquid crystals (LCLCs), a special class of lyotropic LCs comprising rigid aromatic molecules with peripheral hydrophilic ionic and hydrogen-bond groups, form linear aggregates which are held together by non-covalent interactions, leading to self-assembled nematic (N) phase or columnar (M-phase) with a hexagonal arrangement [25, 26]. Common LCLC examples include sunset yellow (SSY), an azo dye used as a food additive, and disodium cromoglycate (DSCG), an anti-asthmatic drug, used in the current study. While the nematic phase consists of short columnar stacks which appear at room temperature for low LCLS concentrations, the columnar phase emerges at high concentrations wherein the stacks self-assemble into two-dimensional hexagonal array of extended columns in the M-phase [27]. In both the N- phase and the M-phase, the director orients parallel to the columnar axis of the stacks [28, 29], with the latter manifesting as archetypal herringbone or spherulite textures [30].

LCLC confined within nanoporous microfluidic devices, as in our PDMS-based microchannels, may undergo isothermal phase transition under appropriate conditions [30]. The presence of nanopores in PDMS drives selective absorption of water molecules (Figure 1c), thus enhancing the concentration and shifting the thermodynamic phase of the DSCG. Yet, how confinement dimensions, and surface properties control the initiation and dynamics of LCLC phase transitions remain unknown. Here, we leverage PDMS-based microfluidics to confine DSCG, and control the phase transitions from disc-shaped mesogens (isotropic) to aligned cylindrical stacks (nematic), and finally into the honeycomb structures of the M-phase. By systematically varying the surface wettability (and anchoring), confinement cross-section and shape, and the channel topography, we program phase transitions at will under isothermal conditions, within generic microfluidic geometries. Supported by a simple mathematical model, our results uncover spatio-temporal evolution of self-assembled LCLC structures, offering a surface-controlled physical pathway for molecular assembly and cargo transport, with potentially far-reaching ramifications on our current understanding of the prebiotic evolution mediated by nanoporous clay materials like montmorillonite [31].

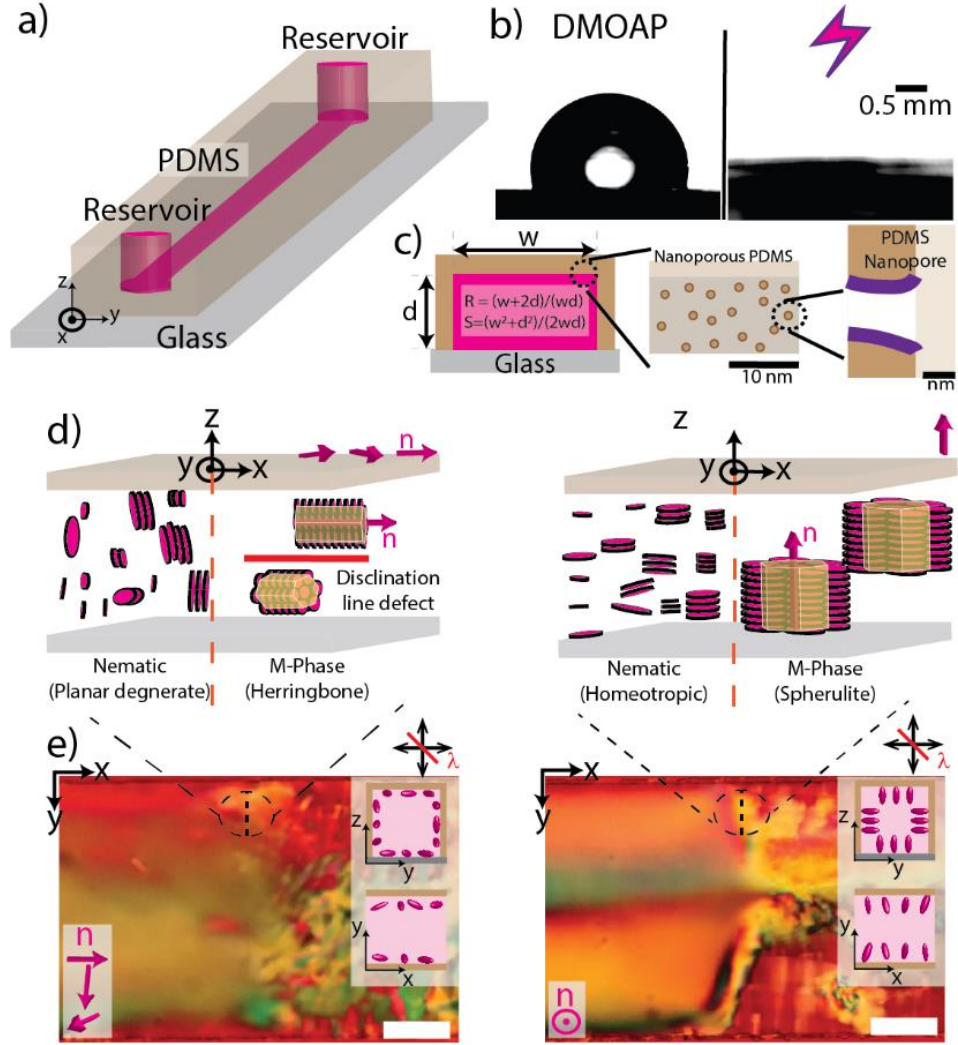

**Figure 1. Nanoporous microfluidic confinements induce lyotropic phase transition.** (**a**) Schematic representation of a glass-PDMS microfluidic channel with both openings acting as reservoirs at the initial LCLC concentrations. The channel depth was kept constant ($d \approx 10$ μm), whereas the width and overall channel geometry were varied in our experiments. (**b**) The wettability of PDMS walls was varied by appropriate surface treatments: coating with the silane DMOAP rendered the surfaces hydrophobic, contact angle of water ≈ $106.90 \pm 2.95$ degrees (n = 14); while exposure to air plasma turned the surfaces hydrophilic, contact angle ≈ 0 degree. The contact angle of water on untreated PDMS surface was found to be $100.60 \pm 0.27$ degrees (n=10). (**c**) Varying the aspect ratio of the channels ($A$ = width /depth = $w/d$) alters the surface-to-volume ratio ($R$) and the shape factor ($S$) of the microfluidic confinements, defined respectively as $R = (w + 2d)/wd$, and $S = (w^2 + d^2)/2wd$. PDMS nanopores, indicated as darker circles, retain the native hydrophobicity even after the DMOAP and plasma surface treatments, thus generating a relative difference in the wettability that drives the transport and selective uptake of water molecules from the LCLC phase. (**d**) Molecular ordering of the LCLC corresponding to the nematic and columnar phases, as observed by polarized optical microscopy (POM) with 530 nm retardation λ-plate The N-M transition front is clearly distinguished by the difference in the texture on either side: nematic and M-phase within plasma treated channels (left panel), and within DMOAP-treated microchannels (right panel); **n** denotes the local director. The vertical dashed lines in (**d**) and (**e**) indicate the regions of transition from nematic to (left image) herringbone and (right image) spherulites texture. Scale bar: 50 μm

DSCG molecules in an aqueous solution exist as disc-shaped mesogens in the isotropic state which gets stacked to form cylindrical aggregates forming a nematic state. With further increase in concentration – driven either by lowering temperature or water content – the cylindrical rod-shaped aggregates assemble into a hexagonal honeycomb structure forming M-phase, while the alignment of these aggregates are still determined by the surface they interact with (Figure 1d). Such change in phase can occur within microfluidic channels when water is take up by PDMS, resulting in the formation of distinct anchoring-specific textures when the M-phase emerges (Figure 1e). It must be noted that cylindrical ports ($\phi$: 1.5mm, $h$: ~6mm) at the either ends of the channels have much larger volumes than the volume of the DSCG confined within the microfluidic chips ($l$: ~10mm, $w$: 50 to 350 μm, $d$: 10 to 25 μm). As the three PDMS-walls selectively absorb water molecules, the volume of water lost from the DSCG solution in the channel is compensated by the relatively higher water content present in the DSCG solution located at the ports which serve as reservoirs of DSCG in nematic phase (or isotropic phase, depending on the initial concentration of the experiment). By symmetry arguments, it can be thus shown that molecular influx of water from the ports converges at the midpoint of linear microchannels [32], resulting in a local maximum of the DSCG concentration near the channel center. This was confirmed in our experiments, and consequently the channel central region was identified as the zone of observation for all measurements.

Once the nematic to M-phase transition initiates at the center of the microchannel, a distinct nematic-M-phase front propagates along either direction, ultimately filling the entire channel length with M-phase. Figure 1e shows a snapshot of the propagating N-M-phase interface. Similarly, when we start our experiments using DSCG in isotropic phase, first the nematic phase appears, followed by the M-phase. Here, we will focus on the appearance of the M-phase texture under distinct anchoring conditions from an initial nematic phase. The grain structures formed by the M-phase is presented in Figure 2. The effect of the surface treatment could be clearly distinguished by quantifying the grain widths for the plasma and DMOAP treated channels. We employ digital image processing techniques to arrive at the texture parameters: first we image the freshly formed M-phase textures between two different cross-polarizer orientations (at 0° and 45° relative to the channel length), and then extract pixel-by-pixel difference in the polarized signals to arrive at the grain sizes presented in Figures 2 a, b. This analysis is conducted across entire width of the channels, to finally obtain the variation of the grain sizes in the transverse direction (Figures 2c and d). Further details on the image analysis algorithm can be found in the Supplementary materials. While homeotropically aligned self-assembled hexagonal structures yielded larger projected grain width, the planar anchored surfaces supported grain sizes with lower mean width (Figures 2c and d), in agreement with existing literature [33]. We further observe that, for a given surface anchoring condition, the grain size increases consistently as one moves from the confining walls toward the long symmetry axis of the microchannel, bearing a qualitative similarity with solidification of liquid metals cast in molds. In freshly formed solid metal phase, grain widths are smaller closer to the surface (known as the chill zone ), which then gradually increase as the distance from the surface increases [34]. Analogous to the rate of water uptake by PDMS surfaces (*i.e.*, mass transport), it is the rate of cooling associated with liquid-solid metal systems that determines the grain dimensions and their coherence [35-37]. The dependence of the grain textures on the rate of the heat transport allows us to hypothesize that the N-M-phase transition can be equivalently tuned by regulating the uptake of water by the PDMS surfaces.

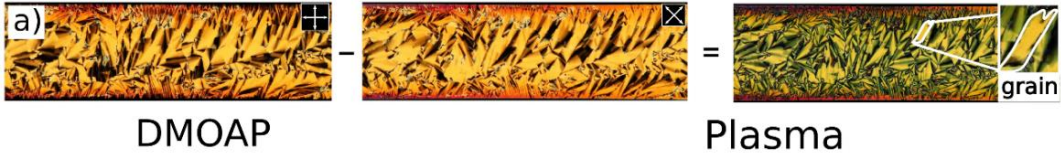
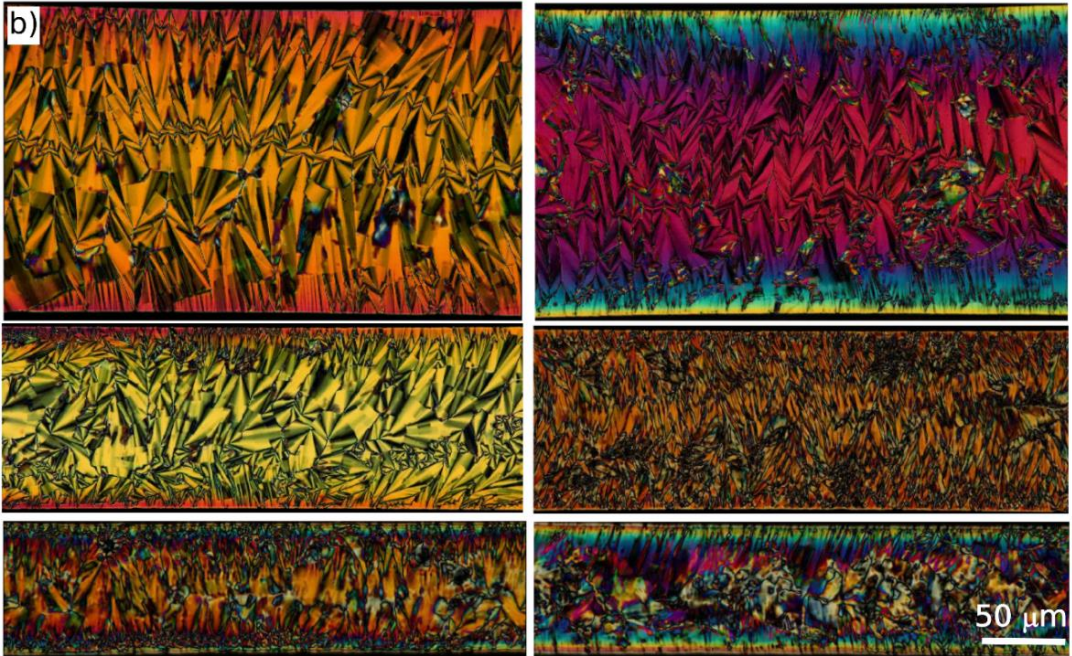
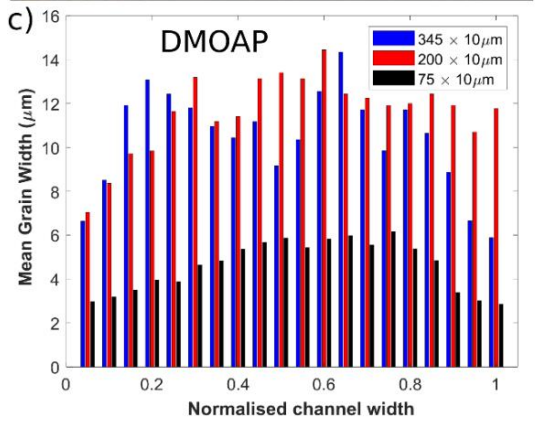
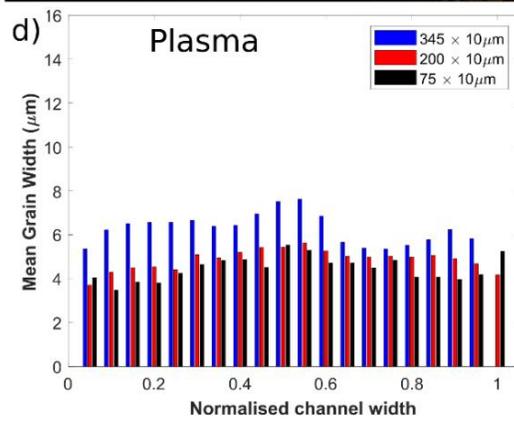
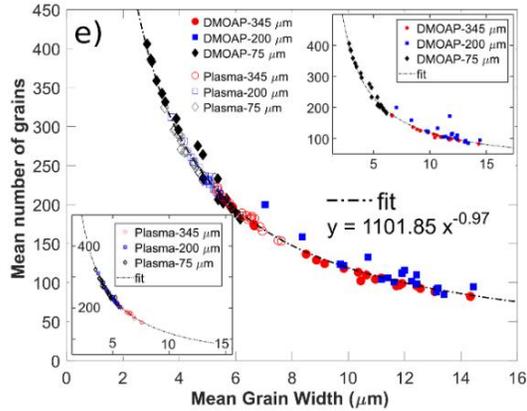
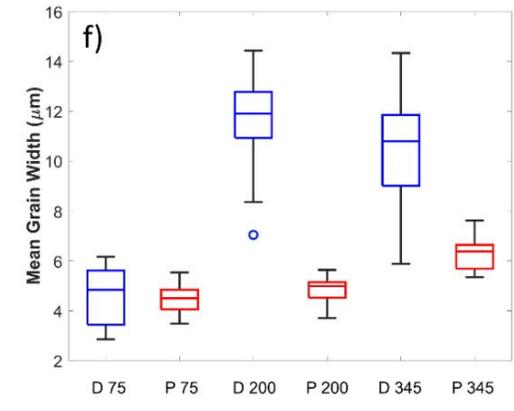

**Figure 2. Columnar (M) phase textures are governed by the microfluidic confinement and surface anchoring.**
**(a)** The M-phase grain boundaries are clearly visualized after quantitative image processing, whereby polarized optical micrographs are obtained for different orientations of crossed-polarizers (at 0 and 45 degrees to the channel length) of a given region. The absolute difference of the polarized light signals reveals the grain sizes of the columnar phase. In contrast, the smooth nematic textures do not show any grainy feature. **(b)** Quantitative image processing reveals M-phase textures for the DMOAP- (left column) and plasma- (right column) treated confinements. The channel dimensions span ($w$ x $d$): 350 µm × 10 µm, 200 × 10 µm and 75 × 10 µm from top to bottom in both columns. **(c)** Overall, the mean grain size is larger for DMOAP-treated channels compared to the plasma-treated confinements **(d)**, due to the relative difference in the projected areas of the spherulite (homeotropic) and herringbone (planar) textures, as shown in Figure 1d. For a given surface-treatment, the grain size scaled with the channel width, with the largest grains observed in the vicinity of the channel center line (at normalized width ~ 0.5). Wider channels yield larger grains and vice versa; the corresponding grain number distributions are provided as supplementary data. **(e)** Grain number and grain width data obtained from microchannels with different surface anchoring conditions and aspect ratios fall on a master curve, $y = ax^b$ where $y$ and $x$ represent the grain number and size respectively, with a = 1101.85 and b = -0.97. Insets present the distribution of the data points on the master curve for the DMOAP- (top) and plasma- (bottom) treated channels. (f) Box plots capture the variation of the mean grain width across the cases discussed above.

In the case of DSCG solution, the transition from the nematic to the M-phase is analogous to liquid-solid metal solidification process, as the fluid-like nematic phase transforms to a solid-like M-phase. As the diffusion of heat during cooling in metal casting promotes the phase transition, in our case the selective diffusion of water molecules into the nanoporous PDMS matrix drives the local enhancement of DSCG concentration, ultimately initiating the N-M phase transition. The effect of anchoring conditions on M-phase formation is clearly captured in Figure 2: on average, the characteristic grain size under homeotropic anchoring conditions is twice larger (Figure 2c) than the grain sizes observed in channels with degenerate planar anchoring conditions (in the middle of the plasma-treated channels channels, the size ranges from 4 µm to 8 µm, Figure 2d). This trend is reversed when we quantify the total number of grains within the same normalized bins (Supplementary data): the density, *i.e.*, number of grains in a given area, maximizes in the vicinity of the channel walls, decreasing gradually toward the central region of the channel. Overall, the number of grains in homeotropic conditions is lower than that in planar conditions. Interestingly, the variation of the grain number over the grain size corresponding over any region falls onto a master curve $y = ax^b$, where $y$ and $x$ represent the grain number and size respectively, with *a* = 1101.85 and *b* = -0.97 (Figure 2e). The corresponding insets present the trend for DMOAP- (top) and plasma- (bottom) treated channels: within DMOAP-treated, the grain size and number vary widely with respect to the channel dimensions; whereas within plasma-treated channels, the impact of channel dimensions is less pronounced. In summary, the effects of anchoring and confinement dimensions set a trade-off, such that the anchoring–dependent difference in the relative grain sizes and numbers get compensated as channel widths are reduced. This is seen clearly for the case of $w = 75$ µm, the mean grain width is statistically comparable under both anchoring conditions (Figure 2f). The aspect ratio of the freshly formed M-phase grains is longer in plasma-treated channels, compared to those formed under DMOAP-treated confinements, manifesting typically as the herringbone and spherulite textures respectively. Since the long axis of the hexagonal assembly of cylindrical stacks of DSCG (see Figure 1d) is aligned parallel to the channel inner surfaces, the corresponding grains – comprising several hexagonal assemblies packed together – develop with a higher aspect ratio. In the case of homeotropic anchoring, the circular projection of the cylindrical stacks aligns parallel the surface (*i.e.*, the long axis is perpendicular to the channel surface), thus resulting into relatively larger grain widths. The decrease in the grain width observed with the reduction of the channel width could be reasoned due to the faster diffusion of water under the larger surface-to-volume ratio, $R = (w+2d)/wd$, where $w$ and $d$ respectively indicate the width and depth of the microchannels.

Building up on Chvorinov's relation for the solidification time in metal casting, given by $t = \beta \left(\frac{V}{\alpha}\right)^{\eta}$, where $V, \alpha, \eta$ are the mold volume, surface area, and a constant; and $\beta$ is the mold constant, a function of the material properties including density and heat capacity [38], we propose that the time required for the nematic to the M-phase transition should be inversely proportional to $R$, the surface-to-volume ratio of the channel. The basic theory behind the metal solidification process has been reasonably well established and it is a phase transition occurring in a single component system. In our study, the phase transition of DSCG is a two-component system where water and DSCG molecules exists. Therefore, it is important to understand the water absorption kinetics by PDMS which eventually triggers the phase transition of nematic to the M-phase. To understand the water absorption kinetics by PDMS, we record the increase in the weight of PDMS rectangular molds (similar size, different anchoring conditions) immersed in bath of de-ionized water, relative to corresponding dry weights. The percentage change (*PC*) of the weight of the PDMS molds pre-treated with plasma and DMOAP, shows a striking difference (Figures 3a and b). While hydrophilic surfaces are known to better absorb water molecules than hydrophobic surfaces [39], here we report an opposite trend: the water absorption capacity (equilibrium *PC*%), and the corresponding rate of absorption by the plasma-treated PDMS molds were lower compared to the DMOAP-treated molds of similar dimensions. As control, the untreated PDMS molds (which are inherently hydrophobic) exhibited better water absorption property (both uptake rate and saturation *PC*) than the plasma-treated molds, however weaker when compared to DMOAP-treated counterparts. Our results indicate that the hydrophilic PDMS surface hinders transport of water molecules through the nanopores (as shown in Figure 3 b), due to the generation of a wettability gradient between the continuous surface (rendered hydrophilic due to plasma-treatment) and the nanopores which retain their intrinsic hydrophobicity. Such a gradient of wettability can cause water molecules to remain strongly associated with the continuous PDMS surface, rather than diffusing into the nanopores [40]. However, the trend reverses when the continuous surface has lower wettability (induced by the DMOAP-treatment), relative to the nanopores. In this case, the water molecules are preferentially driven into the nanopores, thus enhancing the overall water uptake rates.

Different surface-treatments impact not only the anchoring of the DSCG molecules, but in addition, control the phase transition time. In order to validate the suggested mechanism, we varied the temperature of the water bath where the PDMS molds were immersed. Since the surface interaction energy between the pre-treated PDMS surface and water molecules is altered, impeding (or promoting in case of DMOAP-treatment) the diffusion of water molecules into the nanoporous matrix should depend on the thermal energy. Higher thermal energy should enhance the uptake kinetics irrespective of the anchoring effects. This is confirmed in Figure 3b inset, which presents the variation of the experimental rate constant $k_2$, as a function to the bath temperature. The data was fitted using a pseudo-second-order kinetics reaction, and the fitted lines represent the Arrhenius equation $k_2 = A_0 e^{-\frac{E_a}{RT}}$ where $A_0$ is a constant and $E_a$ is the activation energy. For the current set of data, the activation energy offers a quantitative measure of the energy barrier that the water molecules need to overcome, which is higher for the plasma-treated surface ($E_{a,planar}$ ~4.48 x $10^4$ J/mol) than that of the DMOAP-treated surfaces ($E_{a,homeo}$ ~3.54 x $10^4$ J/mol).

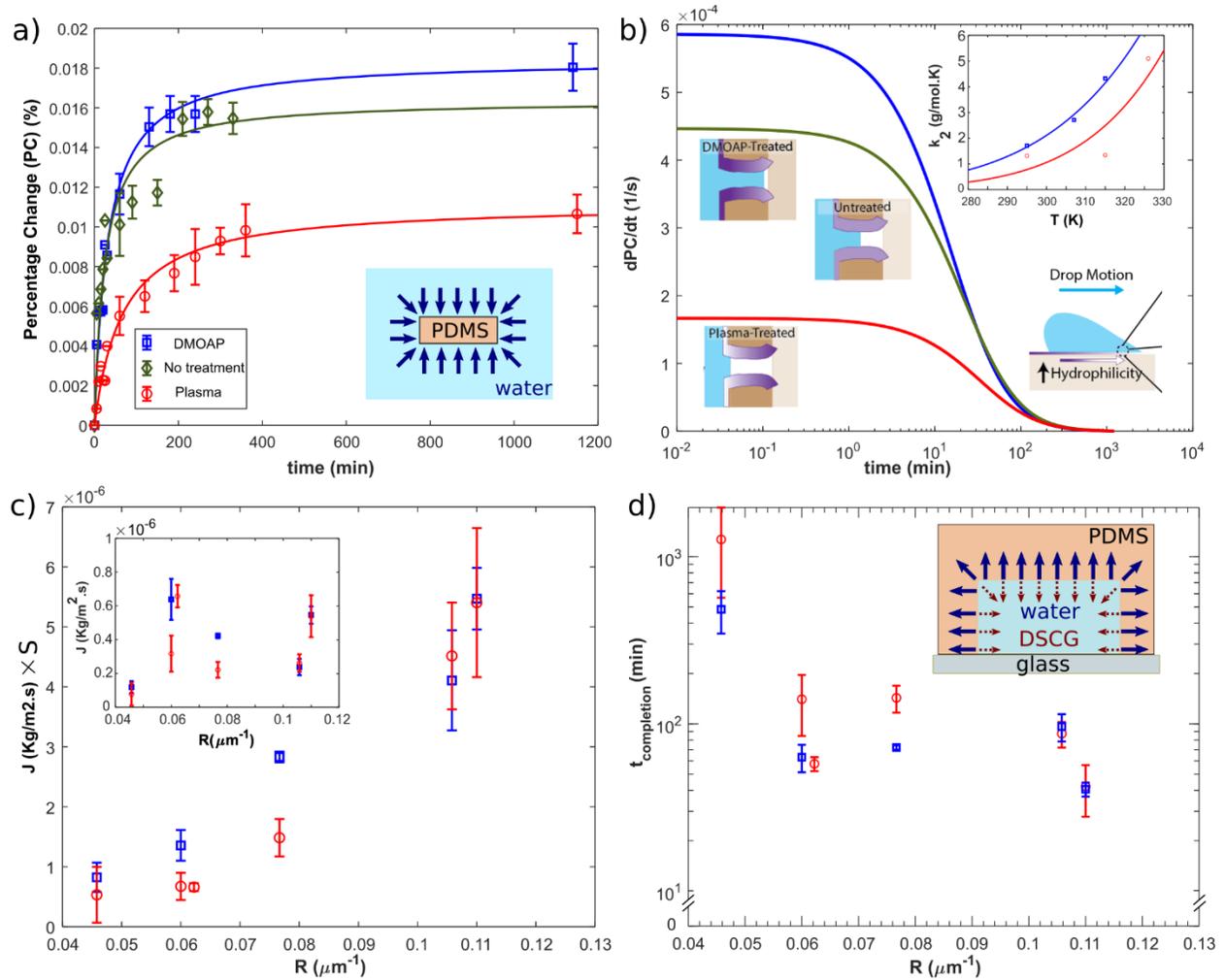

**Figure 3. Surface wettability and confinement cross-section govern the rate isothermal N-M phase transition**. (a) Percentage change (*PC*, in %) in the weight of surface-treated PDMS blocks submersed in water, shown as (blue □) and (red ○) for DMOAP- and plasma-treated surfaces respectively, while the native PDMS surface is indicated by (black ◊). DMOAP enhances PDMS hydrophobicity, while plasma treatment enhances surface hydrophilicity: this leads to a relative reduction (for DMOAP) and enhancement (for plasma) of wettability relative to the native PDMS-nanopore wettability. Solid lines in respective colors fit the experimental data using a pseudo-second order kinetics. The experiments were done at room temperature (23 ± 1° C). The inset shows the schematic of water being absorbed

by the rectangular PDMS blocks. **(b)** The gradient d*PC*/d*t* is obtained from the pseudo second order fit as shown in (a) that is plotted against time (also at 23 ± 1º C). The inset plots the rate constant $k_2$ for *PC* values obtained at three different experimental temperatures: T ≈ 23º C, 34º C, and 42º C for DMOAP-treated samples; T ≈ 23º C, 42º C, and 53º C for the plasma-treated samples. The data is fitted to Arrhenius equation, further information on the fitting is available in the supplementary section. Blue and red lines indicate the DMOAP- and plasma-treated cases respectively. **(c)** The water uptake rate per unit surface area (*J*) is regulated by the cross-sectional shape of the PDMS-glass microchannels: *J* x *S* shows a monotonic increase with the corresponding surface-to-volume ratio (*R*). Here *S* is the shape factor as defined in Figure 1c. The inset shows the variation of *J* with the surface-to-volume ratio, *R*. **(d)** The time taken for the M-phase to cover a given region of observation (initially nematic phase) reduces with the surface-to-volume ratio, for both the wettability conditions. The inset shows schematic of PDMS-glass microchannel cross-section, the water uptake proceeds through the three bounding PDMS surfaces thus leading to increase of local DSCG concentration at the PDMS surfaces.

Figure 3c presents the flux, *J*, of water uptake by the PDMS per unit time and surface area, considering the three PDMS-walls of the rectangular channel. *J* is estimated by the time it takes to complete the phase transition over a given region of observation (Figure 3d), once the channel has been filled with 14% DSCG solution (nematic phase). As discussed earlier, we fix the observation point close to the middle of the channel length, allowing the water molecules absorbed by the PDMS to be replenished by the water molecules from the DSCG reservoirs in nematic phase at the channel ports. Confirming previous tracer experiments by Randall *et al*. [32], we provided a physical framework for the N-M phase transition in LCLCs confined within naoporous microfluidic settings. We infer that the local DSCG concentrations will increase at the middle of the channel, marking the initiation of the N-M phase transition. The inset plots the variation of *J* as a function of the surface-to-volume ratio, *R*, for both the plasma- and the DMOAP-treated channels. The *J* vs *R* plots fail to capture a clear trend, and moreover indicates that for relatively close values of *R* (0.06 µm$^{-1}$ and 0.062 µm$^{-1}$), values of *J* can be significantly different. This seeming discrepancy could be resolved by accounting for the shape of the microfluidic channels, in addition to their surface-to-volume ratios. Empirically, we extracted the shape factor, $S = (w^2 + d^2)/2wd$, where the symbols indicated usual terms (see Table 1). By multiplying the flux *J* with *S*, we recover a clear trend, as shown in the *J* x *S* vs *R* plot in Figure 3c. Interestingly, the shape of casting molds has also been shown to affect the metal solidification time, which however was not predicted by Chvorinov's rule [41].

Upon nucleation of the N-M interface lose to the center of the microchannel, a distinct front travels on either direction, towards the channel ports. Figure 4a presents the N-M front position over time for the two anchoring conditions. The slope of the position vs time plot (inset) gives the front propagation speed, with a typical value around 4 µm/min for plasma-treated confinements, and 2.5 µm/min – 3 µm/min for the DMOAP-mediated front propagation. It is to be note that the front speed decelerates over a distance: the deceleration rate is faster for the plasma-treated case than for the DMOAP-treated channel (Figure 4a, inset). As the water molecules from the nematic phase of the DSCG solution confined within the ports (reservoirs) enter and dilute the relatively higher concentration of the DSCG solution within the channel, the effective PDMS surface area in contact with the nematic phase decreases, resulting in the deceleration of the N-M front speed. Intriguingly, the difference in the plasma- and DMOAP- mediated front speeds contradicts the trends in the water uptake rates reported in Figures 3a, b. As the DMOAP-treated channel has shown to uptake water faster (Figure 3c), one expects the N-M front speed to be faster as well, since the rate of molecular transport should be proportional to the speed of the N-M (as reported in the single-component liquid-solid metal casting process [42]). This contrasting dynamics in the two-component water-DSCG systems is driven potentially by the emergence of high anisotropic elasticity at the phase transition. Higher the *J*, higher is the influx of water molecules from the nematic DSCG solution (in the ports) to the middle of the channels, thus resulting in further dilution and slowing of the N-M front.

Once the M-phase nucleates, the propagating front gradually fills up the entire channel without leaving cracks or physical discontinuities. This is in contrast to our observations using isotropic crystal forming sodium chloride solution (Figures 4b, c), indicating the potential role of elastic forces in driving the phase transitions. Saturated NaCl (30 % *w/v*) solution was clear when filled in channel and after few hours, as expected, the crystal nucleated at the centre. The crystals however, didn't grow continuously as in case of DSCG, and left behind multiple cracks and local discontinuities. The clear contrasting differences in these two materials could be found in Figure 4c. The time sequence captures the differences in the evaporative patterns on of NaCl (top panel) vs DMOAP (lower panel): discrete NaCl crystals appear scattered as the drying droplet recedes over time due to evaporation. The DSCG on the other hand leaves a trail of tightly knitted grain structures as the evaporating drop recedes. We formulate a simple mathmerical model to capture the dynamics of the front propagation in the microfluidic geometry (Figure 4d, see Supplementary section for details of the 1-D model). Using a spring-mass system, we obtain a qualitative match with our experimental data (Figure 4e), and demonstrate that the local front speed can be systematically tuned by varying the spring constant, $k/m$ (a proxy for the emergent elastic interactions at the phase transition) in combination with the volume of the M-phase formed from a given volume of co-existing N-M phase (Ø).

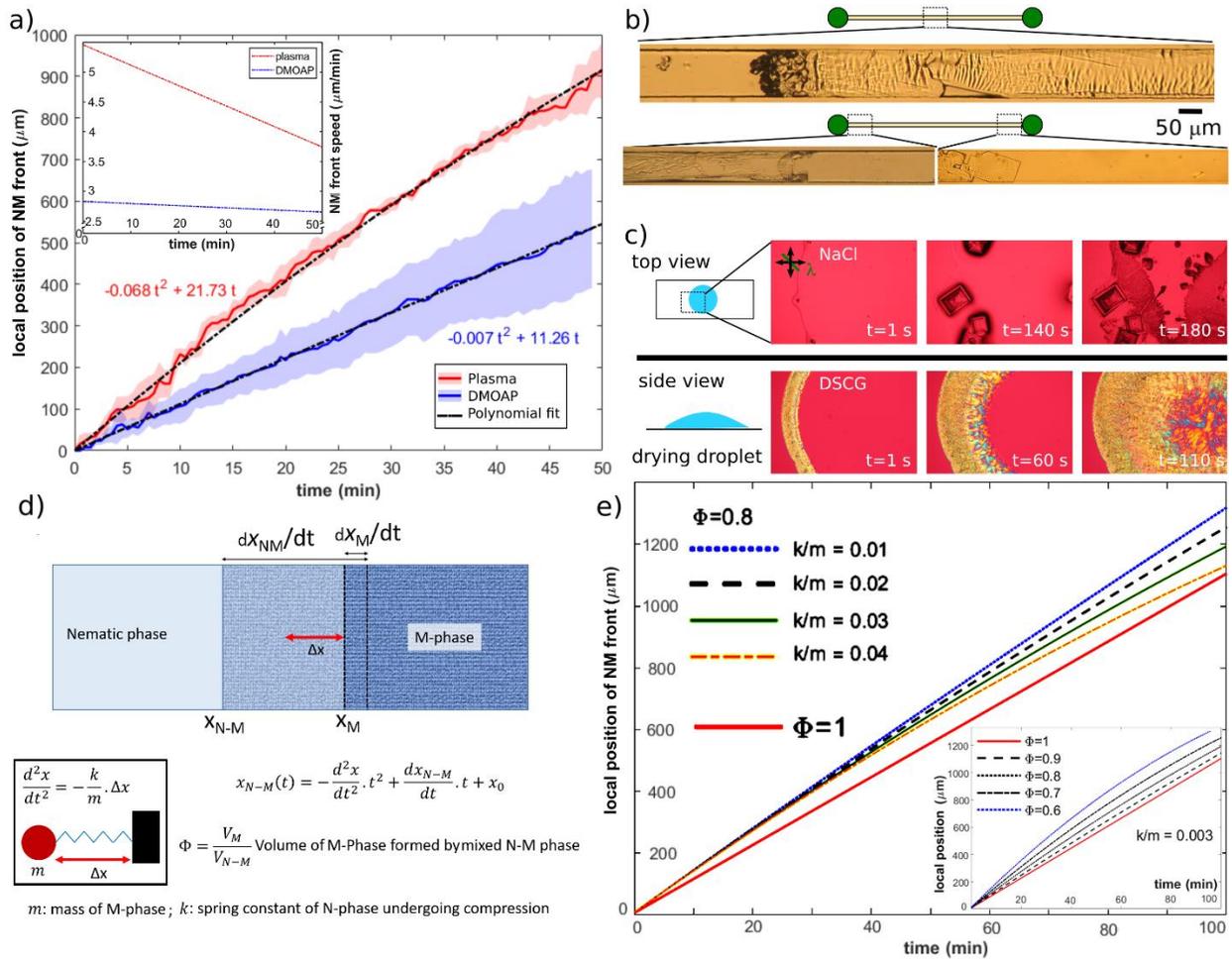

**Figure 4. The nematic-M-phase interface in plasma-treated channels travel faster than that within DMOAP-treated channels**. (**a**) Position of the N-M front over time for the 12% DSCG-filled 200×15 µm channel for plasma (○) and DMOAP (□), obtained over 3 experimental replicates. Inset plots the front speeds over time for the corresponding cases. (**b**) Contrasting textures are observed when 30% (w/v) NaCl is filled in 200×15 µm channel (top panel): the crystal formation occurs similarly at the centre of channel however the emerging textures is discontinuous. (**c**) Discontinuous domains emerge when 30% (w/v) NaCl solution undergoes phase transition (top row image

sequence), in contrast to the 6% DSCG solution drying on a plasma-treated glass slide, showing a continuous texture. The drying droplets are millimetric in dimension. **(d)** Schematic of the mathematical modelling framework, with the key parameters: spring constant, $k$, and $\emptyset$, the volume of compact M-phase emerging from the mixed N-M phase volume. Here, $m$ is the mass of M-phase in the N-M-phase. **(e)** By varying the parameters $k$ and $\emptyset$ systematically, the general trend of the N-M front position was captured.

We leverage the spatial and temporal insights of the nematic to M-phase transition to demonstrate programmability of lyotropic transition under isothermal conditions. As a first case, we consider T-shaped microchannels variable arm lengths, offering a simple but geometrically more complex confinement compared to the straight channels (Figures 5a, b). In case of T-shaped channels with the horizontal arms longer than the vertical arm (Figure 5a), the N-M phase transition initiates simultaneously in both the horizontal arms. The reservoirs at end of each arm drive the transition to M-phase first in the horizontal arms, which then converge towards the center, finally propagating towards the bottom arm (Figure 5a). When we invert the geometrical condition, *i.e.,* the vertical arm is longer that the horizontal ones, the transition to M-phase initiated first within the vertical arm, followed by spreading across the rest of the T-shaped channel (Figure 5b). We extend the programmable N-M phase transition concept to confinements of higher geometric complexity [13] such as the septapod, seven armed channel in Figure 5c. With one arm shorter than the rest, the N-M front nucleates first within the longer arms (arms b – g in Figure 5c), and propagates away from the ports to finally converge at the septapod junction. Thereafter the N-M front can continue travelling to the shorter arm (arm-a), before fully filling up the shorter microfluidic arm (arm-a).

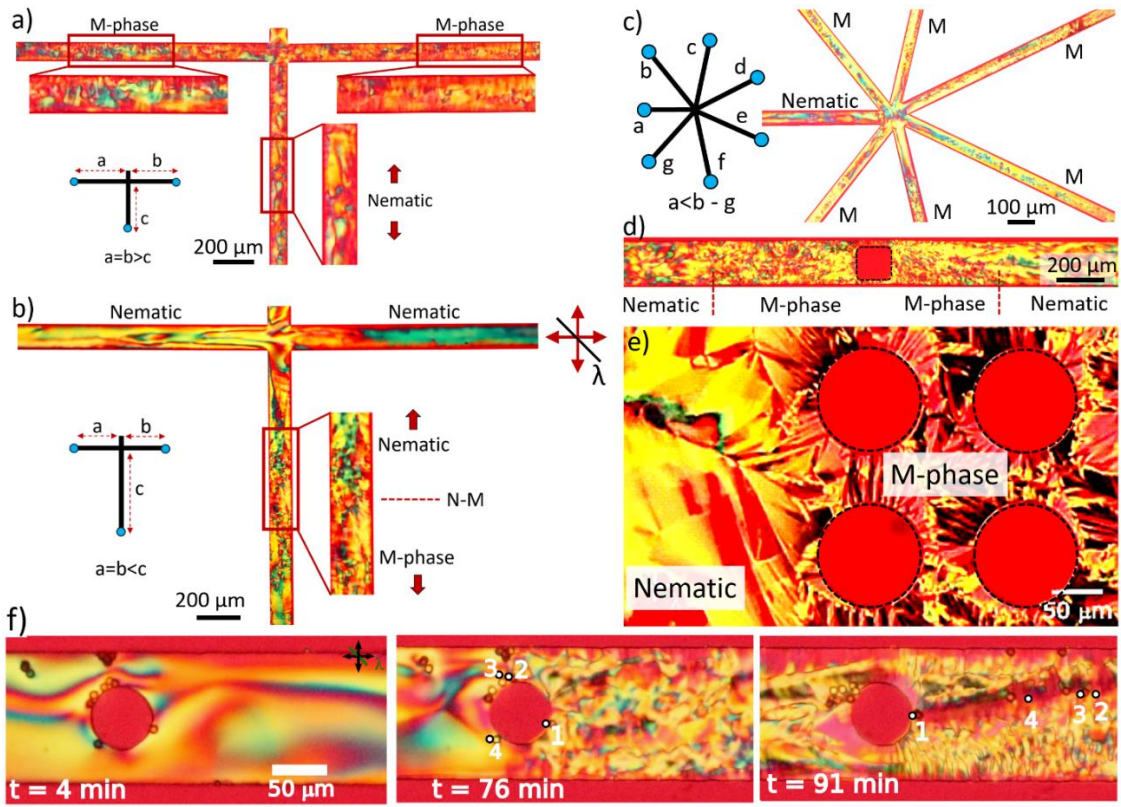

**Figure 5. Spatio-temporal programming of isothermal phase transition in nanoporous microchannels**. (**a**) T-junction channel with 3 arms connected to the ports. By varying the length of one arm, relative to the other two arms, the N-M phase transition nucleation point could be offset. The longer arms (arms a and b) of the T-channel yield M-phase, whereas the shorter arm (arm-c) maintains nematic phase. (**b**) Now by changing the relative arm lengths, N-M phase transition could be triggered in the longer c-arm, while other two arms retained the nematic phase. (**c**) The concept was extended to multi-arm channels, shown here for a septapod where the shortest arm (arm-a) retains the nematic phase. (**d**) Insertion of topographical features can induce local N-M phase transition: close to the square pillar (marked with dotted black lines) M-phase is observed, while far away from the pillar, nematic phase is recorded. (**e**) M-phase localization using a pillar network: Pillar-dense regions undergo phase transition from nematic to M-phase, while pillar-sparse regions retain nematic phase. (**f**) Elasto-advective transport of ~5 µm particles due to emergent long-range order during the N-M transition offers novel spatio-temporal particle manipulation concepts. Typical particle transport rate is $2.1 \pm 0.3$ µm/min.

In addition to channel geometry, we alter the embedded topography by incorporating micro-pillars within the microfluidic devices (Figures 5d, and e). A micropillar, due to its nanoprous surface, can enhance the local water uptake rate, thereby driving the system to an N-M phase transition faster. Figure 5d presents a square-shaped cylindrical PDMS pillar placed in the middle of a straight channel, which has induced the N-M phase transition locally. Such pillars could be offset from the central position, thereby shifting the location and rapidity of the M-phase formation. These results enable flexible lyotropic phase-transition concepts mediated by the confinement geometry and internal topography of the channel surfaces. Finally, by introducing micron sized tracer particles in the channel, we reveal the emergence of elasto-advective flows during the course of the nematic-M-phase transition (Figure 5f). Particles dispersed in the lyotropic LC phase get pulled toward the less mobile M-Phase, opposite to the direction of the travelling N-M front. The inherent elastic nature of LCLC does not allow the M-phase to be discontinuous (unlike in NaCl solution), resulting in stored elastic energy which can be released by infusing water molecules into the system. With *et al.* have utilized multiple arm microfluidic channels with controlled flow rates to spatially induce lyotropic phase transition [43]. Similarly other biopolymers like DNA, and pectin have been made to produce compact aggregates, spatially at desired location in the microfluidic channels, utilizing the precise control of flow properties and diffusion of these molecules along the channels [44, 45]. Our results, in contrast to the previous reports, do not employ any external driving mechanisms like pump-driven flows, and instead is uniquely dependent on the uptake of water by the nanoporous PDMS walls. The high degree of spatio-temporal control – enabled by the geometry, anchoring, wettability and topography – pave the way for producing self-assembled complex functional materials [46]. Going beyond applications of lyotropic phase transition in a range of stimuli-responsive materials relevant for drug delivery and wound healing [47-52], our results offer a physical and programmable route to biologically-relevant phase transitions under generic nanoporous settings, evoking a fresh take on surface-mediated molecular self-assembly in the prebiotic world.

43. With, S. *et al.* Fast diffusion-limited lyotropic phase transitions studied in situ using continuous flow microfluidics/microfocus-SAXS. *Langmuir* 30, 12494–12502, 2014.

44. Iliescu, C. & Tresset, G. Microfluidics-driven strategy for size-controlled DNA compaction by slow diffusion through water stream. *Chemistry of Materials* 27, 8193–8197, 2015.

45. Marquis, M., Davy, J., Fang, A. & Renard, D. Microfluidics-assisted diffusion self-assembly: Toward the control of the shape and size of pectin hydrogel microparticles. *Biomacromolecules* 15, 1568–1578, 2014.

46. Mezzenga, R. *et al.* Nature-Inspired Design and Application of Lipidic Lyotropic Liquid Crystals. *Advanced Materials* 31, 1900818, 2019.

47. Huang, Y. & Gui, S. Factors affecting the structure of lyotropic liquid crystals and the correlation between structure and drug diffusion. *RSC Advances* 8, 6978–6987, 2018.

48. Wang, C. *et al*. Improving Water-Absorption and Mechanical Strength: Lyotropic Liquid Crystalline–Based Spray Dressings as a Candidate Wound Management System. *AAPS Pharm. Sci. Tech.* 23, 1–10, 2022.

49. Zhang, X. *et al.* Smart phase transformation system based on lyotropic liquid crystalline hard capsules for sustained release of hydrophilic and hydrophobic drugs. *Drug Delivery* 27, 449–459, 2020.

50. Zhang, H. & Wang, Z. Phase transition and release kinetics of polyphenols encapsulated lyotropic liquid crystals. *International Journal of Pharmaceutics* 565, 283–293, 2019.

51. Fan, J., Zhang, H., Yi, M., Liu, F. & Wang, Z. Temperature induced phase transformation and in vitro release kinetic study of dihydromyricetin-encapsulated lyotropic liquid crystal. *Journal of Molecular Liquids* 274, 690–698, 2019.

52. Yue, X. et al. A bacteria-resistant and self-healing spray dressing based on lyotropic liquid crystals to treat infected post-operative wounds. *Journal of Materials Chemistry B* 9, 8121–8137, 2021.

53. Sengupta, A. Topological Microfluidics: Nematic Liquid Crystals and Nematic Colloids in Microfluidic Environment. *Springer International Publishing*, Cham, Switzerland, 2013.
22

**Methods**

We use a soft lithography to design the microfluidic devices, and quantify the isotropic-to-nematic-to-columnar phase transitions using polarization optical microscopy. The concentration dependent emergent LCLC textures for different confinement and surface conditions (anchoring and wettability) are compared quantitatively using image processing techniques. Overall, DMOAP-treated surfaces yield homeotropic anchoring and low surface wettability, and plasma-treatment generated planar anchoring and high surface wettability [9, 30]. The intrinsic nanopores of PDMS remain unaffected due to these surface treatments. The dynamics of the nematic-columnar phase transition are captured using timelapse imaging mode under fixed temperature conditions. The time of the appearance of the M-phase – detected by the moving N-M interface – shows a strong dependence on the channel geometry (width and depth) and surface anchoring (degenerate planar vs. homeotropic). The time taken for the entire region of observation to undergo phase transition ($t_{completion}$) was noted for each case, alongside measurement of the N-M front propagation speed. The experiments reported here were repeated a minimum of 3 times to determine the statistics, and to ensure the reproducibility of our results. Table 1 lists the parameters, units and the symbols used in this work.

TABLE 1. Glossary of symbols used in the study

| Parameter | Unit | Symbol |
|---|---|---|
| Number of replicates | - | $n$ |
| Width | m | $w$ |
| Depth | m | $d$ |
| Length | m | $L$ |
| Diameter of cylindrical ports (inlet/outlet) | m | $\phi$ |
| Surface-to-volume ratio | m$^{-1}$ | $R$ |
| Shape factor | 1 | $S$ |
| Contact angle | ° | $\theta$ |
| Temperature | °C, K | $T$ |
| Time | s, min | $t$ |
| Volume | m$^3$ | $V$ |
| Water mass flux | Kg/(m$^2$.s) | $J$ |
| Velocity | m/s | $v$ |
| Concentration | % (w/w) | $c$ |
| Diffusion coefficient | m$^2$/s | D |
| Mass | Kg | $m$ |

| Percentage change | % | *PC* |
| Aspect ratio | 1 | *A* |
| Rate constant | g/(mol.K) | *k₂* |
| Activation energy | J | *Eₐ* |
| Local director | - | **n** |

*Materials and solution preparation*

We prepared the lyotropic chromonic liquid crystal solutions using DSCG in deionized millipore water, vortexed for 30 min and stirred at room temperature for 12 h to obtain a homogeneous solution. The silane, Disodium cromoglycate and dimethyloctadecyl [3-(trimethoxysilyl) propyl] ammonium chloride solution (DMOAP) was purchased from Sigma-Aldrich. 4-cyano-4-pentyl-1 and used as received. All the solvents used were of GC purity ($\geq$ 99.8%) sourced from Carl-Roth.

*Microfluidic confinement*

We prepared the PDMS-glass microfluidic devices using the fabrication protocols outlined previously [23]. Our microfluidic chips comprised 3 walls of polydimethylsiloxane (PDMS, Sylgard 184, Dow Corning) reliefs, prepared by following the standard soft lithography techniques. The PDMS reliefs were bonded to glass substrates via air plasma exposure. Prior to bonding, the glass substrates were thoroughly cleaned with isopropanol and dried on a hot plate (80 °C for approximately 30 min). Microchannels with rectangular cross-sections having different geometries are considered in this work: linear, T-shaped, septapod, and channels with topographical features (single or regular array of pillars). For the linear channels, the depth ($d$) kept was kept at 10 μm, while three different channel widths ($w$) were investigated: 75 μm, 200 μm, and 345 μm. The distance between the inlet and the outlet port was typically set to 20 mm, defining the length ($l$) of the channel. Cylindrical holes (1.5 mm diameter) at the ends of the channels provided the housing for microfluidic tubings, which also served as the reservoirs.

*Filling the microfluidic channels filled with lyotropic LC*

The experimental protocols are based on our previous publication [30]. Solutions with 12 or 14 wt. % of DSCG solution was gradually filled in microfluidic channels using syringe pump by placing the glass side on hot plate at 45° C to make sure the solution is in isotropic state, thus preventing flow induced surface alignment. The open ends of the Teflon tubings, fixed on both ends (ports) of the channel, were sealed with playdough to ensure no evaporative losses during the long-term timelapse image acquisition. Before starting the experiments, the midpoint of the channel was marked on the glass side (using a marker), and used as reference region of interest for the POM. All measurements were carried out around the central region of the channel, close to the geometric midpoint, allowing us to capture the initiation of the nematic to M-phase transition.

*Surface anchoring and boundary conditions*

Microfluidic devices were functionalized for contrasting wettability and surface anchoring conditions: air plasma exposure yielded high wettability (low contact angle) and degenerate planar anchoring condition; whereas DMOAP-treatment resulted in low wettability (high contact angle) and homeotropic surface anchoring. Since the effect of plasma-treatment degrades under normal environmental conditions [9], freshly treated channels were immediately filled with the LCLC. To induced homeotropic alignment, the channels were first filled with 0.4 wt.% aqueous solution of DMOAP, and then thermally cured at 80 °C for 2 h, rendering both the PDMS and glass surfaces hydrophobic with homeotropic boundary conditions. The surface wettability, quantified in terms of the static contact angle of de-ionized water on the functionalized surfaces, was measured using goniometer. Detailed steps of the plasma exposure and DMOAP treatment can be found in refs [12]. The surface wettability and anchoring conditions obtained using above treatments were respectively validated using de-ionized water and thermotropic nematic LC, 4-cyano-4-pentyl-1, 10-biphenyl (5CB, procured from Synthon Chemicals and used as received) as outlined in our previous studies [53].

*Polarization optical microscopy and image acquisition*

The LCLC director orientation, phase transition, and the elasto-advective transport of microscale particles in the PDMS-glass microfluidic chips were studied using a polarized light microscope (Eclipse LV100N-POL, Nikon), equipped with different objectives ×5, ×10, ×20, and ×40 objectives. The sample temperature was maintained with a Linkam (PE120) heating/cooling stage. Experiments were conducted under different orientations of the crossed-polarizers using transmitted white light: (a) the microfluidic channel parallel to one of the crossed-polarizers, (b) the channels oriented at 45° with respect to the polarizers, and (c) with and without the insertion of a 530 nm phase-retardation plate. POM images (6000 × 4000 resolution), and full HD coloured videos (1920 × 1080 resolution) recorded at a frame rate of 25 frames per second were acquired using a digital Canon EOS77D camera. The acquired POM data were quantified using a combination of the ImageJ, VideoMach, and MATLAB softwares [13, 21].

*Bulk water uptake by PDMS molds*

The intrinsic nanoporous structure of PDMS allows it to uptake liquids depending on their polarity, including water and liquid crystals [53]. In this work, the rate of water uptake by bulk PDMS was obtained by measuring the difference in weight of PDMS molds before and after immersing them in a bath of de-ionized water. PDMS molds of dimension 40 mm x 20 mm x 6 mm were cut from larger chunks, cleaned with ethanol in an ultrasound bath, and dried on hot plate before recording the dry weight of the PDMS molds. The dry weight of the molds were measured using a sensitive balance (XSR204DR, Mettler Toledo, BENELUX) with a precision of 100 µg, right after the molds were treated using DMOAP and air plasma following the protocols mentioned above. Thereafter the molds were immersed fully in water baths maintained at a fixed temperature. Similar experiments were carried out using untreated PDMS molds, providing control measurements. The immersed molds were weighed periodically to determine the change of the weight due to uptake of water. Similar experiments were conducted for three different temperatures in order to obtain the pseudo-second-order kinetics. Details of the data fitting can be found in the supplementary section.

*Contact angle measurements*

The contact angle of water (de-ionized Millipore) on the surface-treated substrates were measured using a standard goniometer (Dataphysics, OCA 15-EC, Filderstadt, Germany) in static mode via sessile drop technique, with a precision of ± 0.1 degrees. A small but defined volume (tens of microliters) of water was dispensed on the treated substrate, which was then imaged live using an inbuilt software (SCA20). This allowed us to identify the base line (liquid-solid interface) and the contour of the water drop (liquid-air interface). The contact angle was obtained by fitting the contour line with appropriate equation provided by the goniometer software. It is important to note that the volume of the dispensed water should be kept low in order to ensure accurate wettability measurement. Large volumes of sessile drops could create spurious measurements due to gravity-driven spreading of the droplets on the substrates of interest.

# Supplementary Materials for

# Spatio-temporal programming of lyotropic phase transition in nanoporous microfluidic confinements

Vamseekrishna Ulaganathan[1] and Anupam Sengupta[1*]

[1]Physics of Living Matter Group, Department of Physics and Materials Science, University of Luxembourg, 162 A, Avenue de la Faencerie, L-1511, Luxembourg City, Luxembourg.

* To whom correspondence may be addressed: anupam.sengupta@uni.lu

**This PDF file includes (6 pages):**
Supplementary Figures: 4 (Fig. S1 to S4)
Supplementary movies: 2
Supplementary References: 2

## 1. Grain size analysis

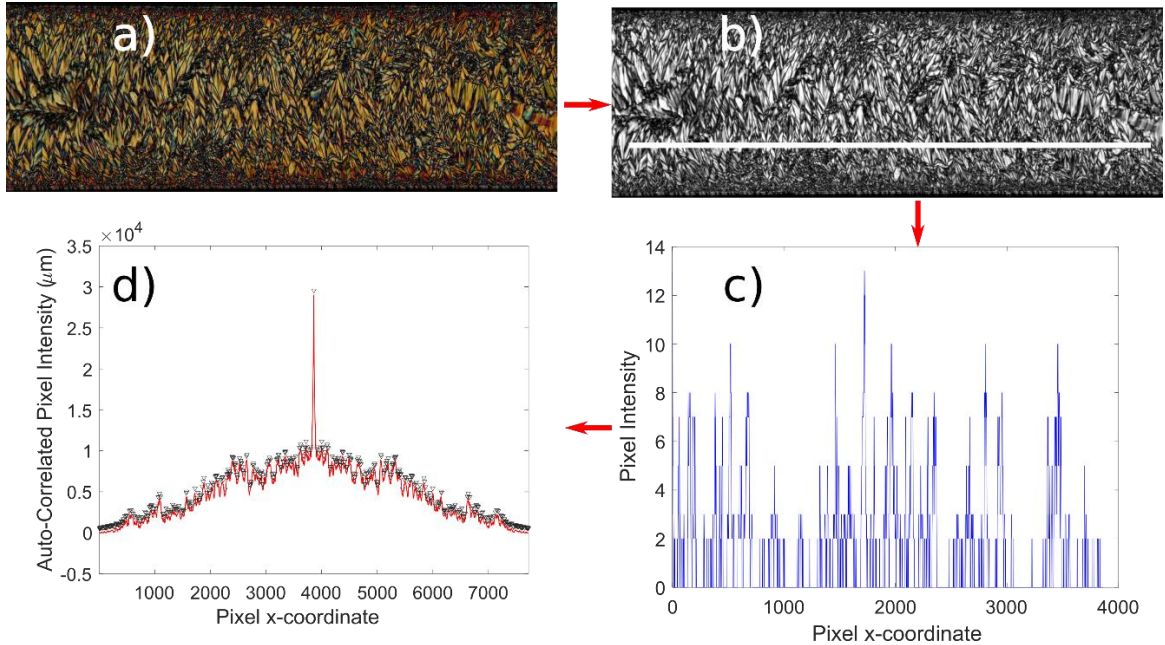

Figure S1. Image analysis pipeline, performed on MATLAB for estimating the grain size of M-phase. a) absolute difference image between cross-polarized images at 0° and 45°. b) converted to gray scale image after applying gaussian smooth function. The white line shows the pixels that were scanned along the x-axis. c) the pixel intensity plotted against the x-coordinates. d) the pixel intensities in the (c) were cross correlation against each other and plotted against x-coordinates. The (▼) symbols mark the local peaks recognised.

The grain size analysis presented in Figure 2 (in main manuscript text) is performed using the method shown in Figure S1. The absolute difference image shown in Figure S1(a) comprises of the raw data containing the grain features. We extracted the information from these features, i.e., the grain width (as most grains are oriented perpendicular to the x-axis), by scanning the pixel intensities of grey scale image (Figure S1(b)). We note here that employing edge detection for these images led to inconsistencies in the grain detection, hence this method was not considered beyond a few basic trials. By extracting the pixel intensities along the line on x-axis and cross-correlating it we obtained the peaks, revealing the brightest region between the dark edges of the grain (around the midpoint of the grain boundaries). Therefore, the difference in the distance between these peaks give us an estimate of the grain width. This analysis gives us a semi-quantitative understanding of the grain size distribution as function of proximity to the channel borders (as discussed in the main manuscript text).

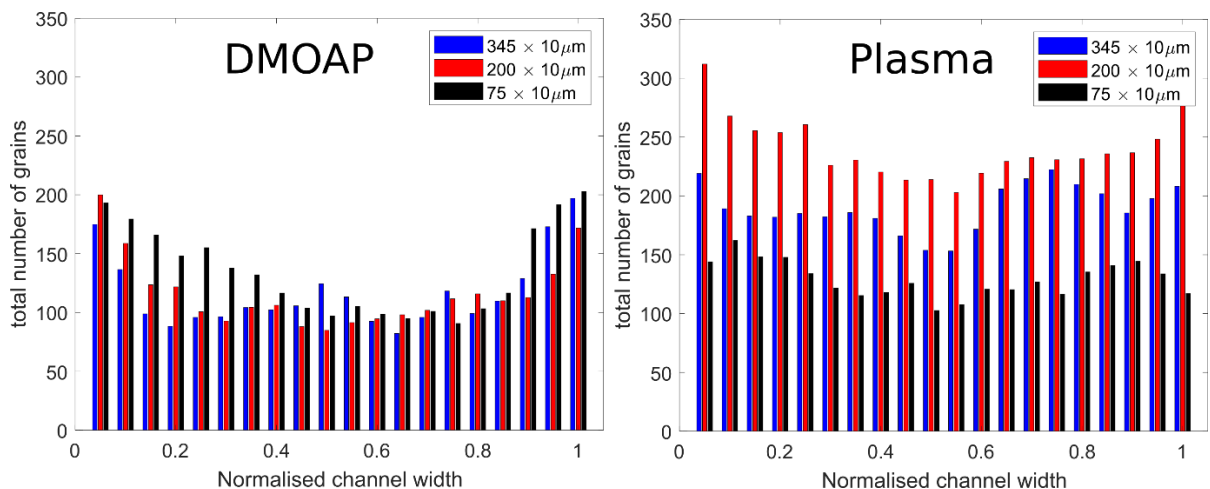

Figure S2. The total number of grains as a functions of normalised channel width for the data presented in Figure 2 a & b for both DMOAP and plasma treated channels.

As shown in Figure 2 c & d (of the main text), the grain widths are smaller at the edges and the larger at the centre of the channels, resulting in higher number of grains at the channel boundaries compared to the centre. The influence of the channel surface treatment (leading to different anchoring conditions) on the grain size distribution has been discussed in the main text.

## 2. Calculation of water uptake rate in microfluidic channels

In this section we elaborate how the $J$ values presented in Figure 3 (of the main manuscript text) were calculated. The initial concentration of DSCG solution filled in the microfluidic channels was 14% DSCG. At this concentration, the solution is in the nematic state at room temperature (~21°C). The threshold concentration at which the DSCG solution reaches M-phase at 21°C is 28% DSCG. We attribute the change in concentration of DSCG (visualized as the textural changes due to the phase transition) to the uptake of water by the PDMS surfaces. By taking the difference between the time at which the channels were filled ($t_0$) and the time at which the M-phase transition is completed within the region of interest ($t_f$), gives us the time of completion of phase transition, $t_{completion}$ (presented in Figure 3D). The length of the channel in the given observation frame, L is 588 μm (seen under 10X magnification), and by knowing the channel width and depth, we extract the volume. So, the volume of water, $\Delta V$ that needs to be taken up by the PDMS surfaces to trigger a concentration change from 14% to 28% for the given channel dimensions, was estimated. We normalized the water uptake to the available surface area (SA), i.e., the 3 PDMS sides of the channel walls. The water uptake rate, $J$ is given as $J = \frac{\Delta V \times \rho}{SA \times t_{completion}}$ (kg/m²s).

## 3. Water uptake by PDMS

Adsorption is a process where the molecules of a substance bind to specific sites on a surface of another material and its kinetics is usually predicted by a pseudo-first-order kinetic model or pseudo-second-order kinetic model, among others. [1,2]

In case of water uptake by the PDMS the process is absorption rather than adsorption. But since it is analogous, (absorption being bulk process while adsorption being surface limited process) these models might be utilized. Here we chose pseudo-second-order kinetic model since it gave us the best fit (see Figure S3).

The differential equation for **pseudo-second-order kinetic model** is given as follows.

$$\frac{dq_t}{dt} = k_2(q_e - q_t)^2 \qquad (3)$$

From which the following form is obtained.

$$t/q_t = 1/{k_2 q_e^2} + t/q_e \qquad (4)$$

By plotting $t/q_t$ Vs. $t$ we extract the $q_e$ and $k_2$.
The adsorption process depends on the concentration of adsorbing species which is usually solute molecules in a solvent adsorbing on some solid surface [1,2]. In our case water concentration is practically infinite (since the volume of water available is much greater than maximum amount that can be absorbed by the PDMS chunks). Therefore, the absorption or water uptake should be a function of pores concentration or sites where water could penetrate into the PDMS in a given surface area. So the amount of water that saturates given volume (or mass) of PDMS should be function of the concentration of it pores per unit surface area.

Since water uptake could be a function of the concentration of pores/sites, P on the PDMS surface, pseudo-first-order reaction is utilized assuming the rate of water uptake is linearly proportional to P.

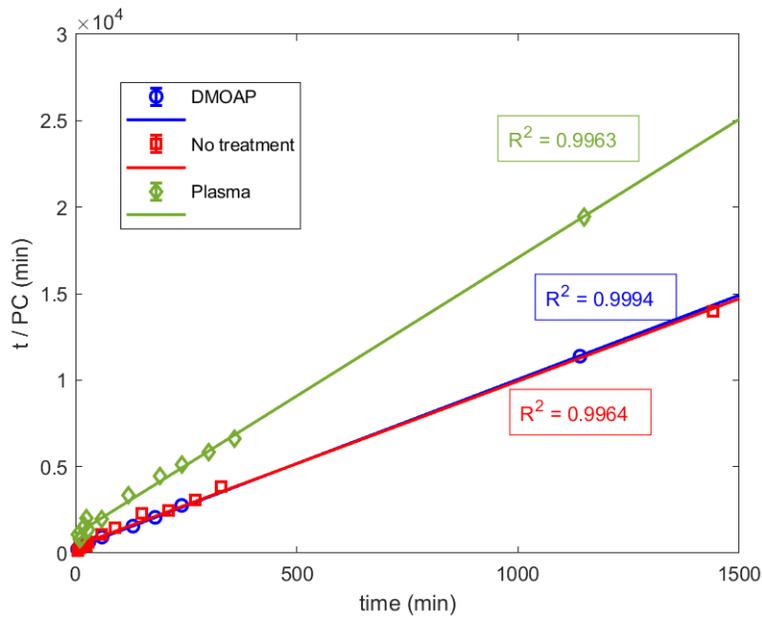

**Figure S3**. t/PC vs. time plot showing the goodness of the fit.

## 4. One-dimensional (1-D) simulation

To understand the phase transition of the N-M phase in microfluidic channels and the elasticity emerging during this transition, a 1-dimensional mass attached to the spring is simulated. Following Hooke's law, $F = -Kx$, where $F$ is the force acting on the mass attached to the spring, K is the spring constant and $x$ is the length of spring extension or compression. A visualization of the spring model in action can be seen in the supplementary movie file *Movie_1D-Model_phase transition_stable.mp4*.

The following are the parameters chosen for the simulation. Mass = 1, rest length (minimum length where the spring is not under tension) = 0.1, Spring constant =20, initial length=1. The initial setup of the simulation consists of 20 objects of mass =1 that were connected with springs with the parameters mentioned above. The springs are under stretched condition and they experience extensile and contractile forces from each other as they are compressing. This is analogous to a condition where DSCG molecules are concentrating as the space between them is reducing as the water is being taken up by the PDMS. At a distance of 0.4 between two objects, we introduce a condition where the objects and the interconnecting spring become static, resembling the phase transition to M-Phase. The initial formation of static objects gives directionality to the rest of the objects connected to springs as they compress towards the initially formed static objects. The emerging elastic reorganization of the system gives rise to elasto-advective transport of microscale particles dispersed in the system (Fig. S4).

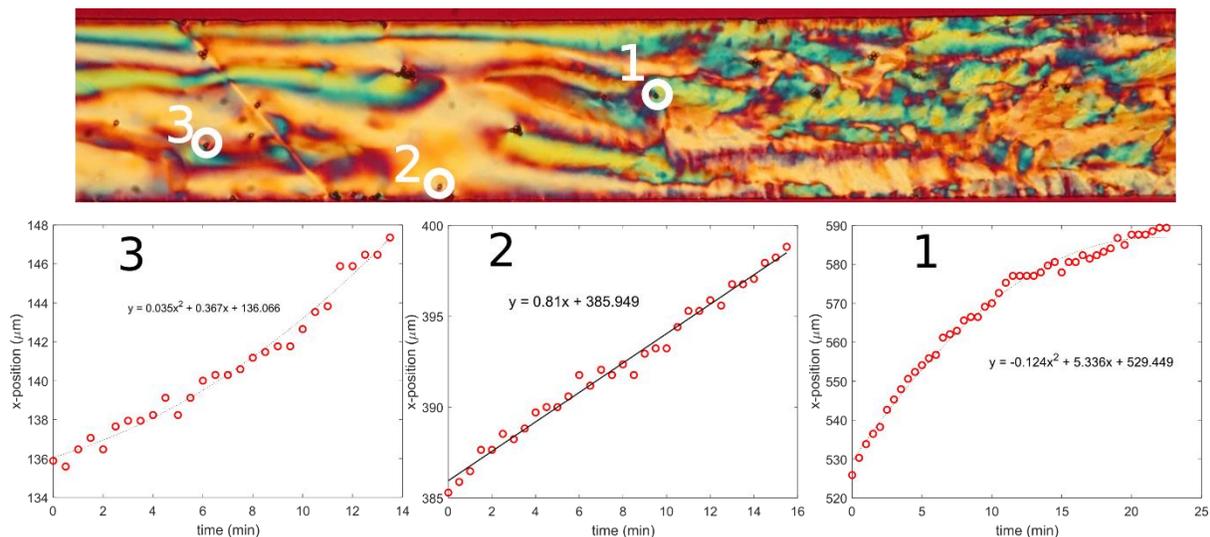

**Figure S4.** Particle speeds due to emergent elasticity arising during M-phase transition. The corresponding movie can be found as supplementary movie file (*Movie_Particle Speeds.mp4*).